\documentclass[twocolumn,showpacs,preprintnumbers,nofootinbib,amsmath,amssymb]{revtex4}
\usepackage{graphicx}
\usepackage{dcolumn}
\usepackage{bm}
\usepackage{amscd}

\newcommand\half{{\frac{1}{2}}}
\def\nuc#1#2{\relax\ifmmode{}^{#1}{\protect\text{#2}}\else${}^{#1}$#2\fi}
\begin{document}

\title{Continuum discretization methods in a
 composite-particle
scattering  off a nucleus: the benchmark calculations}

\author {O.A. Rubtsova}
\email{rubtsova-olga@yandex.ru}
 \author { V.I. Kukulin}
 \email{kukulin@nucl-th.sinp.msu.ru}
\affiliation{%
Institute of Nuclear Physics, Moscow State University, 119991
Moscow, Russia}
\author{A.M. Moro}
\email{moro@us.es}
 \affiliation{Departamento de FAMN, Universidad de Sevilla, Apartado 1065. E-41080 Sevilla, Spain}
\date{\today}
\begin{abstract}
The direct comparison of two different continuum discretization
methods towards the solution of a composite particle scattering off a
nucleus is presented. The first approach
-- the Continumm-Discretized Coupled Channel method -- is based on
the differential equation formalism, while the second one -- the
Wave-Packet Continuum Discretization method -- uses the integral
equation formulation for the composite-particle scattering problem.
As benchmark calculations we have chosen  the deuteron off a
\nuc{58}{Ni}
 target scattering (as a realistic illustrative example) at three
different incident energies: high, middle and low. Clear
non-vanishing effects of closed inelastic channels at small and
intermediate energies are established.  The elastic cross sections
found in both approaches are  very close to each other for all three
considered energies.
\end{abstract}
\maketitle
\section{Introduction}
Historically the main progress in  studying  the elastic and
inelastic (breakup) composite projectile scattering off a heavy
target was done within the Continuum-Discretized Coupled-Channel
(CDCC) method [1-17] in the framework of the Schroedinger
coupled-channel scheme. In the CDCC approach the total three-body
scattering wave function (in initial versions of the method the
projectile was assumed to be a two-fragment nucleus) is expanded in
a complete basis of the fragments relative motion of the
projectile and the unknown three-dimensional elastic and breakup
channel wave-functions of the projectile c.m.\ motion are obtained by solving 
a set coupled-channel equations.
 This approach has been successfully applied in the field of nuclear reactions
 to find the scattering amplitudes of many particular processes 
for various types of projectiles, like deuterons, $^6$Li,
$^6$He, $^7$Li, $^{12}$C etc.  However, a strong discussion about
the validity of the method arose \cite{Sawada,Austern,Red} because
there were some unclear issues related to  the treatment of the total wave
function breakup components in the asymptotic region.  So, to check
the accuracy and  reliability of  the results attained within this
approach, the convergence of the elastic and breakup amplitudes
(with variation of the momentum bin widths and their number) was
studied in detail  for various approximation schemes \cite{P}. This
study has demonstrated clearly that the approach leads to convergent
results for any reasonable choice of the discretization scheme.
 In
particular, it was shown in Ref.~\cite{P} (for the case of deuteron
scattering off a $^{58}$Ni target as a  test case) that the employment
of a finite number of {\em exact} two-body scattering wave functions
as a basis for the internal projectile motion with energies fixed at
midpoints of the bins (the {\em midpoint} method) leads unexpectedly in
general to slower convergence for the elastic and breakup
amplitudes than the usage of the two-body wave functions averaged
inside  energy bins (the {\em average} method). This result 
seems rather unnatural because the use of averaged wave
functions (i.e.\ wave-packets) as a basis for the internal projectile
motion leads to a wrong asymptotic behavior of the coupling
potentials and scattering wave functions in breakup channels, while
the employment of the {\em exact} wave functions of the projectile
internal subHamiltonian  seems from the first glance to be more
adequate.

Thus, the energy averaging method for  the projectile internal
continuum wave-functions seems to assist to get faster
convergence in the CDCC calculations. The averaging leads to
normalized (i.e.\ of $L_2$-type) wave-functions $\phi_i({\bf r})$, depending on 
the relative coordinate between the two fragments.
At the same time, the motion along the second,
 projectile center of mass, coordinate ${\bf R}$ is considered in
the CDCC approach without any averaging on energy. So, the
three-body wave function for the $i$-th excited channel has an
outgoing asymptotic behavior
$$
\begin{CD}
\Psi({\bf r},{\bf R})@>> {R\to\infty\atop r\ll R}> \phi_i({\bf
r})\frac{\exp({\rm i}K_iR)}{R}. \end{CD}
$$
 Therefore above the
three-body breakup threshold the total wave function $\Psi({\bf
r},{\bf R})$  vanishes at $r\to \infty$ (similarly to the
bound-state function) while the dependence of the total scattering
wave function upon the projectile center of mass coordinate  has the
pure outgoing wave asymptotic behavior. Such an asymmetry for the
motions along the two coordinates at energies  above the breakup
threshold looks not fully justified \cite{Sawada,Austern,Red}
because in this energy region the proper asymptotic behavior over
{\em all spatial variables} might be important.

However, in  case of the  scattering of a weakly bound projectile  
(e.g.\ deuteron or $^7$Li) at
relatively high incident energies ($E\gtrsim 40$ MeV) the motion
along the projectile c.m.\ coordinate ${\bf R}$ proceeds much more
faster than the motion along the relative coordinate ${\bf r}$ (if the
projectile c.m.\ energy is much larger than its binding energy). In
this case  the breakup of the projectile takes place mostly outside the nuclear field of 
the target. In other words,  the projectile is first
excited into its internal-continuum states  and then (already outside
the reaction region) the excited projectile breaks up into its
fragments. Within this picture, the breakup of the projectile in
the target field can be treated as a particular case of inelastic
scattering into the projectile continuum states. Then a proper
averaging over energy bins for the fragments motion in the
discretized continuum gives just a discrete representation for an
exact spectral density function of this subsystem. As a result,
this discretized density function enters  all the elastic and
breakup amplitudes.
 Thus, the
applicability condition for the treatment of the above processes
within the CDCC method should be as follows: $E_{\rm total} \gg
E_{\rm binding}$. However when the projectile kinetic energy $E_{\rm
total}$ is comparable or even a bit larger than its inner binding
energy the other asymptotic channels should be taken into
consideration and thus, the coupled-channel Schroedinger equation
formalism should be replaced by the respective Faddeev approach.

Thus, within its scope of validity, the CDCC approach can be
considered as a rather useful and practical working method to handle
with the realistic problems of composite projectile scattering off a
target. However  it would be very important to test carefully the
accuracy and predictive power of the CDCC results in some {\em
fully independent} way  based on a different approach. The
present paper is devoted just to a careful comparison between the
CDCC and one  alternative method for some typical problem
in the field.

The alternative approach which we have chosen for our study is a
  new integral equation approach based on {\em a total} continuum
discretization \cite{K1,K2,K3,K4} which has  been developed by two
of the present authors some time ago. The approach is based on the
Wave-Packet Continuum Discretization (WPCD) technique which allows
to convert the three-body integral equations into the respective matrix
equation with the matrix elements easily calculable in an analytical
way.  In the WPCD method the motion along both radial coordinates
$\bf r$ and $\bf R$ is considered symmetrically, i.e.\ the respective
averaged wave functions (stationary wave packets) vanishing at
infinity along {\em  both coordinates} are used. Such a fully
discretized integral approach can be used because in the expression for the
elastic scattering amplitude the total
three-body resolvent operator $G(E)$  is bracketed from the left and from the
right by fast decaying factors over all the spatial coordinates:
$$
A_{\rm el}(E)=\langle
\phi_0,\psi_C^{(-)}(E-\epsilon_0)|\bar{V}+{\bar V}G(E){\bar
V}|\phi_0,\psi_C^{(+)}(E-\epsilon_0)\rangle,
$$
where $|\phi_0,\psi_C^{(\pm)}(E-\epsilon_0)\rangle$ is the initial
(final) projectile three-body wave function, so that the "external"
interaction, ${\bar V}$, decaying along the c.m.\ coordinate $R$ and
 the projectile bound-state wave functions $\phi_0({\bf r})$
decaying along the internal coordinate $r$ cuts effectively the
asymptotic parts of the total resolvent $G(E)$. Thus, instead of the
exact three-body resolvent operator $G(E)$, its respective
wave-packet finite-dimensional approximation $\hat{G}(E)$  can be
favorably used \cite{K2,K4}. Therefore, in this integral wave-packet
approach we do not need any asymptotic parts of scattering wave
functions. All that is required is just the inner parts of the wave
functions but with the proper normalization which can be well
approximated via the stationary wave packets in the WPCD approach.
It should be stressed that the effective cutoff of peripheric
three-body resolvent  parts does not depend on the total energy.
Hence, if to neglect the stripping channels (similarly to the
conventional CDCC method) and to treat the elastic  and breakup
channels only, the wave-packet method looks fully justified at low
and high energies as well.

Such a  'totally discretized' integral technique has several evident
advantages as compared to other formulations of the same problem. In
addition to a convenient matrix scheme for a calculation of both the
elastic and breakup amplitudes, it also allows to construct in an
explicit form an optical (energy-dependent and non-local) effective
potential for the projectile--target interaction  which accounts for
 the intermediate inelastic processes { properly} \cite{K2,K4}. Another
distinctive feature of the integral approach is a proper and
straightforward inclusion of closed channels which are treated in
this approach on the same footing as the open channels (see especially
the section III).

So, to make the comparison between two approaches maximally broad we
considered  the
 deuteron elastic scattering by \nuc{58}{Ni} at three
 different incident energies:  high, intermediate and low, where closed channel
 contributions are expected to be quite different. This reaction has been
 also used to compare the CDCC and Faddeev approaches in a recent work \cite{Deltuva}, showing 
very good agreement between both methods. It
 will be vividly demonstrated that the contribution of the closed
 channels, which are commonly neglected in CDCC
 calculations  \cite{J,R1,A,R2,F,L,C1,C2,P,C4},
  raises essentially when the projectile energy
 gets low.  In principle, closed channels can be incorporated within
the CDCC scheme, however, in this case 
the solution of the coupled equations becomes numerically
 unstable.  In this situation, it
is more convenient to solve the problem
using
 the R-matrix method \cite{Th2}. This is an equivalent
way of solving the coupled  equations, but with the advantage
of being more stable numerically as compared to
the traditional radial stepping methods.  However, this reformulation demands
 more computational resources than the conventional coupled-channel
 scheme \cite{Th}. Thus, we made a comprehensive comparison for both
 open and closed channel contributions between the R-matrix CDCC
 and WPCD approaches at different energies.

  The paper is organized as follows. In
 Section II a brief description of the CDCC and WPCD approaches is presented.
 Section III contains  numerical results obtained with these
 two approaches and the discussion is
 placed in Section IV.
 Finally, the summary of the paper is given in Section V.

\section{Brief description of the CDCC and WPCD approaches}
The Hamiltonian  describing the scattering of the two-fragment composite
particle $\{bc\}$ off the target nucleus $A$ is taken in the form:
\begin{equation}
\label{ham}  H=h_{bc}({\bf r})+h_C({\bf R})+V_{b-A}({E_{\rm
lab}}/2)+V_{c-A}({E_{\rm lab}}/2)+\Delta V_C,
\end{equation}
where $h_{bc}$ is the subHamiltonian of the internal $b-c$ motion
(acting on the fragment relative coordinate $\bf r$), $h_C$
--- the subHamiltonian of the projectile center of mass
asymptotic motion including long-range point-like Coulomb
interaction (acting along the c.m.\ coordinate $\bf R$), $V_{c-A}$
and $V_{b-A}$ are energy dependent optical potentials for the
fragment--target system taken at half the incident particle
energy $E_{\rm lab}$ (for the nearly equal mass fragments). Finally,
$\Delta V_C$ is the additional short-range Coulomb interaction
(acting along the c.m.\ coordinate) which is caused by the finite
charge radius of the target nucleus (due to charge distribution in
the target nucleus). To obtain the elastic scattering and breakup
amplitudes for the above scattering problem with the three-body
Hamiltonian (\ref{ham}) one has
 to solve either the three-body Schroedinger equation with proper
asymptotic boundary conditions or the system of   the respective
three-body Lippmann--Schwinger equations.

\subsection{Continuum discretized coupled channel method (a differential equation
formulation).}

In the CDCC approach the continuous spectrum of the $h_{bc}$
subHamiltonian is discretized by dividing the continuous momentum
distribution into a finite number of non-overlapping bins
$[k_{i-1},k_i]_{i=1}^N$ with corresponding averaged (within each
bin) continuum wave functions $\{|\phi_i\rangle\}_{i=1}^N$ and
respective averaged 'channel eigenenergy' values
$\epsilon_i^*\equiv\langle \phi_i|h_{bc}|\phi_i\rangle$. To simplify the 
notation we omit here the angular variables  but they are assumed to
be included. Then, the total three-body wave function of the
Hamiltonian $H$ is expanded over the set of averaged wave functions
describing the projectile internal spectrum (including the
projectile bound state $|\phi_0\rangle$ with the binding energy
$\epsilon_0^*$) :
\begin{equation}
 |\Psi(E)\rangle=\sum_{i=0}^N |\phi_i,\chi_i\rangle,
 \label{Cwf}
\end{equation}
where $\langle {\bf R}|\chi_i\rangle\equiv \chi_i({\bf R})$ are the
 channel wave functions which define the elastic
$(i=0)$ and breakup ($i\neq 0$) amplitudes. The bin wave functions
$\{ |\phi_i\rangle \}_{i=1}^N$ are typically constructed by
averaging the {\em true} continuum states, $\phi(k)$ within the bin
interval:
\begin{equation}
\label{binwf}
|\phi_i\rangle=\frac{1}{\sqrt{\Delta_i}}\int_{k_{i-1}}^{k_i}f(k) |\phi(k)\rangle
{\rm d}k,\quad i=1,\ldots, N
\end{equation}
where $\Delta_i$ is a normalization constant and $f(k)$ is a weight function.
For non-resonant continuum the common choice is $f(k)\equiv 1$, in which
case $\Delta_i=k_i - k_{i-1}$.
Applying the expansion (\ref{Cwf}) to the initial Schroedinger
equation for the Hamiltonian (\ref{ham}) one gets a
system of coupled differential equations for the unknown functions
$\chi_i({\bf R})$:
\begin{equation}
\label{CDCCeq} [h_C({\bf R})+\Delta V_C({\bf R})+V_{ii}({\bf R}) -
(E-\epsilon_i)]\chi_i({\bf R})=\sum_{i'=0\atop i'\neq
i}^{N}V_{ii'}\chi_{i'}({\bf R }),
\end{equation}
where $V_{ii'}({\bf
R})\equiv\langle\phi_i|V_{b-A}+V_{c-A}|\phi_{i'}\rangle$\footnote{The
integral is taken  over the relative $\bf r$ coordinate only.} are
coupling potentials. This system is solved 
with the following boundary conditions \cite{P}:
\begin{equation}
\label{asympt} \chi_i(R){\sim}
I({K}_i,R)\delta_{i0}-\sqrt{\frac{{K}_i} {{K}_0}}S_{i,0}O({K}_i,R),
\end{equation}
where $I$ and $O$ are the Coulomb incoming and outgoing waves, $K_i$
is the c.m. wave number which is related to the 'channel
eigenenergy'
 $\epsilon_i^*$ by energy conservation, i.e.\ 
 $E=\epsilon_i^*+\frac{\hbar^2 K_i^2}{2M}$  (here $M$ is deuteron-target 
 reduced  mass)
  and $S_{i,0}$ are $S$-matrix elements
 for the elastic $(i=0)$ and inelastic $(i\neq 0)$
 scattering\footnote{ For closed channels  ($\epsilon_i>E$) the
 boundary condition should be defined with an
 exponentially decaying and rising functions. In the standard CDCC
  approach the inclusion of
 closed channels causes numerical instabilities 
  due to the unavoidable appearance of exponentially rising asymptotic parts,
   so that the closed channels
 are
 usually not taken into account in these calculations. To
 avoid this problem  the combination of the CDCC and R-matrix  approaches
  can be used.}.

\subsection{Wave-packet continuum discretization method}

In the WPCD approach one applies an analogous discretization of the
continuous spectrum of the $h_{bc}$ subHamiltonian  but making use of {\em energy} bins:
$[\epsilon_{i-1},\epsilon_i]_{i=1}^N$\footnote{It is possible to
make just the same energy and momentum bins in the WPCD and CDCC
approaches using the interrelation
$\epsilon_i=\frac{\hbar^2k_i^2}{2m}$, where $m$ is the
reduced mass of the two projectile constituents.}. In a similar way, one defines a set of stationary
wave packets (WP) as integrals of the exact scattering wave
functions $|\phi(E)\rangle$ of the $h_{bc}$ subHamiltonian over the
respective energy bins:
\begin{equation}
|Z_i\rangle=\frac1{\sqrt{D_i}}\int_{\epsilon_{i-1}}^{\epsilon_i}|\phi(E)\rangle
{\rm d}E,\quad i=1,\ldots, N
\end{equation}
where $D_i\equiv \epsilon_i-\epsilon_{i-1}$ are the bin widths and
the corresponding 'channel eigenenergies' are just the bin
midpoints: $\epsilon_i^*=\half(\epsilon_{i-1}+\epsilon_i)$. The set
of these WP states should be supplemented  with the bound state
$|Z_0\rangle\equiv|\phi_0\rangle$. The main difference in this
discretization procedure with respect to the conventional CDCC one is that we
use here {\em a full set} of WP states which includes both open
and {\em closed} channels.

Besides that we build  a partition of the continuous energy spectrum
of the Coulomb  subHamiltonian $h_C$ acting along the projectile
c.m.\ coordinate $\bf R$ and employ the corresponding set of Coulomb
stationary wave packets (CPs) constructed by averaging the regular
Coulomb continuum wave functions $|\psi^C(E)\rangle$ over energy
bins $[{\cal E}_{j-1},{\cal E}_j]_{j=1}^M$:
\begin{equation}
|X_j^C\rangle=\frac1{\sqrt{\Delta_j}}\int_{{\cal E}_{j-1}}^{{\cal
E}_j}|\psi^C(E)\rangle {\rm d}E,
\end{equation}
where $\Delta_j\equiv {\cal E}_{j}-{\cal E}_{j-1}$ are the bin
widths. Finally, we construct the three-body wave-packet (TWP) basis
set for the elastic  channel asymptotic Hamiltonian,
\begin{equation}
H_{bc}=h_{bc}\oplus h_C,
\end{equation}
as the direct product of the states $|Z_i\rangle$ and
$|X_j^C\rangle$ (including all the required partial wave couplings):
\begin{equation}
|S_{ij}\rangle\equiv|Z_i,X_j\rangle, \quad i=0,\ldots,N;\quad
j=1,\ldots,M.
\end{equation}
Such a three-body packet basis (which is also of the $L_2$ type) is
very convenient for the subsequent investigations. One of its main
advantages is that the matrix of the channel resolvent
$G_{bc}\equiv[E+{\rm i}0 - H_{bc}]^{-1}$ is diagonal in this basis.
Moreover,  the corresponding matrix elements of the three-body
resolvent operator  have an explicit analytical form \cite{K2,K4}
and they depend on the spectrum discretization parameters (i.e.
$\epsilon_i$ and ${\cal E}_j$ values) only.

Due to the fact that the interactions between the projectile
constituents and  the target are given by complex optical-model potentials
the contribution of the intermediate rearrangement (viz.\ stripping)
channels to the elastic scattering or breakup should be  of minor
importance\footnote{Due to the imaginary potentials there are no
bound states in the rearrangement channels, and also the kinetic
energy of the projectile is assumed to be relatively high.}. Therefore, in order to
treat the projectile elastic scattering in the leading order, one
can take into account only those intermediate breakup states whose
asymptotic behavior are   defined by the three-body channel
Hamiltonian $H_{bc}$\footnote{This is just the same assumption as in
the CDCC approach.}.
 So,  the total three-body scattering wave function $|\Psi(E)\rangle$ in this case
  can be found
from a single Lippmann--Schwinger (LS) equation:
\begin{equation}
\label{psils}
|\Psi(E)\rangle=|\Psi_d(E)\rangle+G_{bc}(E)\bar{V}_{bc}|\Psi(E)\rangle,
\end{equation}
where  $\bar{V}_{bc}\equiv V_{b-A}+V_{c-A}+\Delta V_C$ is the
"external" short-range interaction  and
$|\Psi_d(E)\rangle\equiv|Z_0,\psi^C(E-\epsilon_0^*)\rangle$ is the
outgoing wave function for the projectile in its ground state. The long-range Coulomb
part for the point-like Coulomb interaction between the projectile
and target is included into the $G_{bc}$ operator. Next, we make the
projection of Eq.~(\ref{psils}) onto the wave-packet subspace and
expand the total wave function in the three-body wave-packet basis
for the Hamiltonian $H_{bc}$:
\begin{equation}
|\Psi(E)\rangle\approx|\hat{\Psi}(E)\rangle=\sum_{ij}C_{ij}|S_{ij}\rangle.
\end{equation}
Then Eq.~(\ref{psils}) can be rewritten in a
finite-dimensional form
\begin{equation}
\label{fdeq} |\hat{\Psi}(E)\rangle=|S_{0j_0}\rangle
+\hat{G}_{bc}(E)\hat{V}|\hat{\Psi}(E)\rangle,
\end{equation}
where $|S_{0j_0}\rangle\equiv |Z_0,X^C_{j_0}\rangle$ is the packet
state corresponding to the outgoing  wave function $|\Psi_d\rangle$
(the index $j_0$ is defined by the energy rule: $E\in[{\cal
E}_{j_0-1}+\epsilon_0^*,{\cal E}_{j_0}+\epsilon_0^*]$). The
interaction  operator $\hat{V}$ with a hat  is the wave packet
projection of the external interaction $\bar{V}_{bc}$
\begin{equation}
\hat{V}\equiv \sum_{ij,i'j'}|S_{i'j'} \rangle \langle S_{ij}|
V_{ij,i'j'},
\end{equation}
whose  matrix
 elements  in the TWP basis
can be
 interpreted as generalized coupling potentials:
\begin{equation}
V_{ij,i'j'}\equiv \langle S_{ij}|V_{b-A}+V_{c-A}+\Delta
V_C|S_{i'j'}\rangle
\end{equation}
 and may  be calculated either analytically or
numerically. Thus, in this approach  solving the three-body LS
equation is reduced to the solution of the simple matrix equation
(\ref{fdeq})\footnote{It should be stressed that the matrix equation
(\ref{fdeq}) in our case is strongly distinguished from that which
can be deduced from a direct matrix reduction of the three-body LS
equation using some quadrature mesh-points because all the matrix
elements in our case have been averaged over the energy bins
corresponding to  both the active coordinates.}.

 The composite-particle elastic
scattering amplitude can now be written  formally as the following
diagonal (on-shell) matrix element:
\begin{equation}
\label{lamp} A_{\rm el}\approx \frac{[({{\bf V}}^{-1}-{\bf
G}_{bc})^{-1}]_{0j_0,0j_0}}{\Delta_{j_0}},
\end{equation}
where matrices of the respective operators in three-body WP basis
are denoted by the bold letters,
 while  $\Delta_{j_0}$ is the bin width of the $h_C$ continuum partition.
 Moreover, the breakup amplitude can  be readily found as a
non-diagonal (off-shell) matrix element of {\em the same matrix}
$[{\bf V}^{-1}-{\bf G}_{bc}]^{-1}$ in the WP basis \cite{K2,K4}. The
above matrix is essentially a finite-dimensional analog of the exact
transition operator matrix.

Here it is appropriate to summarize the main distinctive features
which distinguish the wave-packet approach from the conventional
CDCC method:
\begin{itemize}
\item[(i)]{The wave-packet basis for the  projectile
 bound and continuum internal states is chosen to go up to high
excitation energies, so that not only open-channel but also
closed-channel contributions are fully included into the WP
three-body calculations.}
\item[(ii)]{The 'internal' and 'external' motions of the projectile
 are taken into consideration in the WP technique on 'equal
footing', i.e.\ using the bases of wave packets averaged over the
respective energy bins for the motions along {\em  both
independent radial coordinates}. So that  the total three-body
resolvent in the transition matrix elements is cut effectively at
large distances by the interaction potentials and also by the bound-state
functions of the projectile in the initial and final states.}
\item[(iii)]{The above mentioned cutoff of the total three-body resolvent
is intimately related  to the integral formulation of the WP
approach, in which a proper asymptotic behavior of coupling
potentials and also an explicit matching procedure between inner and
asymptotic wave functions are not required.}
\item[(iv)]{The kernel of the respective matrix equation in the WP approach
includes rather smoothed (on energy) matrix elements due to the
integration over energy bins for the motions along  all active
coordinates. This important feature of WP matrix kernels can
readily be extended to the  solution of the general
Faddeev equations using the WP-technique \cite{K2}.}
\item[(v)]{The integral formulation of the three- and few-body
scattering problem makes it possible to reformulate easily the
initial problem in terms of the Feshbach projection operator
technique \cite{K4}. This reformulation allows  to reduce the
elastic scattering of a composite projectile to a simple potential
scattering by a non-local and energy-dependent  potential and
moreover to construct this potential in an explicit analytical
form.}
\end{itemize}
So, it would be very instructive to make a direct comparison between
the WPCD and the conventional CDCC results for some standard test
problem.
\section{Comparison between the CDCC and WPCD: the benchmark calculations
 at different incident energies}
As a general numerical test  of the two methods we have chosen the
popular  case of  the elastic scattering of the deuteron off
$^{58}$Ni target at three different incident energies, namely, $E_{\rm
lab}=80,$ 21.6 and 12 MeV. These detailed calculations allow not
only to make more wide comparison between two approaches but also to
deduce from this comparison the role of the closed channels 
at different energies.

For the sake of simplicity, the  n-p interaction is assumed  here to have the 
Gaussian form \cite{R2}:
\begin{equation}
V(r)=-V_0{\rm e}^{-\beta r^2}, \quad V_0=66.99\mbox{ MeV},\quad
\beta=0.415\mbox{ fm}^{-2}.
\end{equation}
Also, keeping in mind the comparison purposes between the two alternative
methods,  we restrict ourselves here to $s$-waves for the n-p relative 
motion \footnote{Higher n-p partial waves would be certainly
important to compare the theoretical results with the respective
experimental data.}.

The nucleon-nucleus interactions are represented by local
optical-model potentials of Woods--Saxon type. In particular, 
we employ here the potentials from the Koning--Delaroche global fits
 \cite{Koning} evaluated at half of the deuteron incident energy. The  mass values
  $m_n=1.0087$ amu (for the neutron),
  $m_p =1.0078$ amu (for the proton) and $m_{\rm Ni}=57.9353$ amu
  (for the target) are adopted here.
\subsection{Choice for parameter values of the momentum distribution in the CDCC
method} Following the standard procedure, in the CDCC calculations
the continuum was truncated by defining a maximum excitation
$E_\mathrm{max}$, and divided into $N$ energy bins, evenly spaced in
the linear momentum. For each bin, a representative wave function is
constructed according to Eq.~(\ref{binwf}) with $f(k)=1$. The number
of bins $N$ was determined in order to achieve convergence of the
elastic angular distribution. For the two lower energies,
$E_\mathrm{lab}=12$~MeV and $E_\mathrm{lab}=21.6$~MeV, the
calculations include both open and  closed channels. Thus,  the
continuum was represented by a set of $N_\mathrm{open}$ bins  up the
maximum available c.m.\ energy, and by $N_\mathrm{closed}$ bins from
this energy to $E_\mathrm{max}$, and hence $N=N_\mathrm{open} +
N_\mathrm{closed}$.

\subsection{Choice for parameter values of the wave-packet bases for the WPCD calculations}
\begin{figure}
\includegraphics[width=9cm]{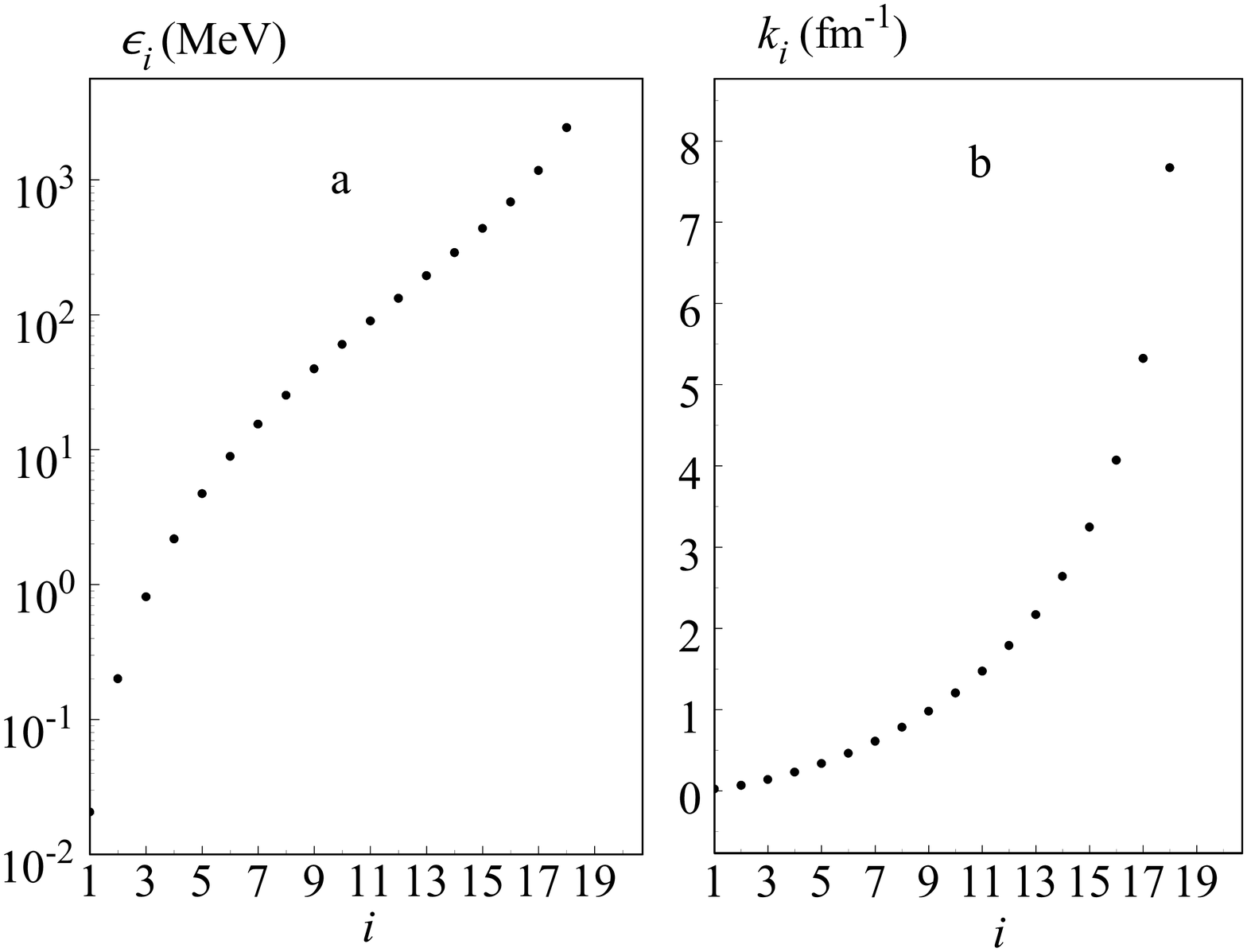}
\caption{\label{fig1} Energy  $\epsilon_i$ (a) and momentum $k_i$
(b) distributions for the n-p continuum discretization constructed
with a Gaussian basis on the generalized Tchebyshev grid
(\ref{grid}). }
\end{figure}

To simplify the construction of the n-p relative motion WPs
$|Z_i\rangle$ and to get simultaneously the bound state  wave
function $|Z_0\rangle$ we apply here a single diagonalization
procedure for the $h_{bc}$ subHamiltonian
 on  the simple  Gaussian basis $|\psi_n\rangle_{n=1}^{N_G}$, i.e. we assume that the
 n-p wave packets are expanded as follows:
 \begin{equation}
 \label{gaub}
 |Z_i\rangle \approx\sum_{n=1}^{N_{G}} A_n |\psi_n\rangle, \quad
 \psi_n(r)=B_n{\rm e}^{-\alpha_nr^2},
\end{equation}
where $B_n$ are normalization factors.
The scale parameters $\alpha_n$ are  defined on the generalized
Tchebyshev grid
\begin{equation}
\label{grid}
\alpha_n=\alpha_0\left[\tan\left(\frac{\pi(2n-1)}{4N_G}\right)\right]^t,\quad
n=1,\ldots,N_G,
\end{equation}
where $\alpha_0$ is the common scale parameter and  $t$ defines  the
distribution sparseness for the Gaussian 'frequencies' $\alpha_n$.
As $t$ become larger the mesh points  $\alpha_n$ are simultaneously
extended to lower and higher values, so that  more longer-range and
shorter-range Gaussian components  appear in the basis.
  The
 energy distribution $\epsilon_i^*$ (and the corresponding
momenta $k_i=\sqrt{\frac{2m\epsilon_i^*}{\hbar^2}}$) obtained with
such a basis occurred to be highly non-homogeneous (i.e.\
non-equidistant), but it covers practically all the energy spectrum
and provides well converged results for the scattering amplitudes
\cite{K1}. Figure \ref{fig1} shows the energy and corresponding
momentum distributions for the following parameter values chosen for the grid: 
$N_G=20$, $t=2$, $\alpha_0=0.1$ fm$^{-2}$.
In practical calculations we found that it is quite sufficient to
include  only the first $N$ basis functions with maximum bin energy
$\epsilon_N$ less than some chosen maximal energy value $E_{\rm
max}$  of the included closed channels. For example, it is clear
from  Fig.~\ref{fig1} that a basis containing the first $N=13$ functions is 
enough for $E_{\rm max}=100$ MeV.

As for the WP basis to describe  the n-p pair c.m.\ motion we employ
(for all  required partial waves $L$) a set of the exact Coulomb
packets of the dimension $M$ constructed also on the Tchebyshev grid
with $t=1$, which corresponds to the  energy distribution:
\begin{equation}
{\cal E}_j=E_0\tan\left(\frac{\pi(2j-1)}{4M}\right),\quad
j=1,\ldots,M,
\end{equation}
where ${\cal E}_j$ are the end points of the energy bins (${\cal
E}_0=0$) and $E_0$ is the incident  energy of the projectile
in the c.m.\ frame.

 \subsection{Results of calculations} 
In this section we will present the particular results for the 
differential elastic cross sections calculated at different incident energies.

\subsubsection{$E_{\rm lab}=$80 MeV}
\begin{figure}
\includegraphics[width=9cm]{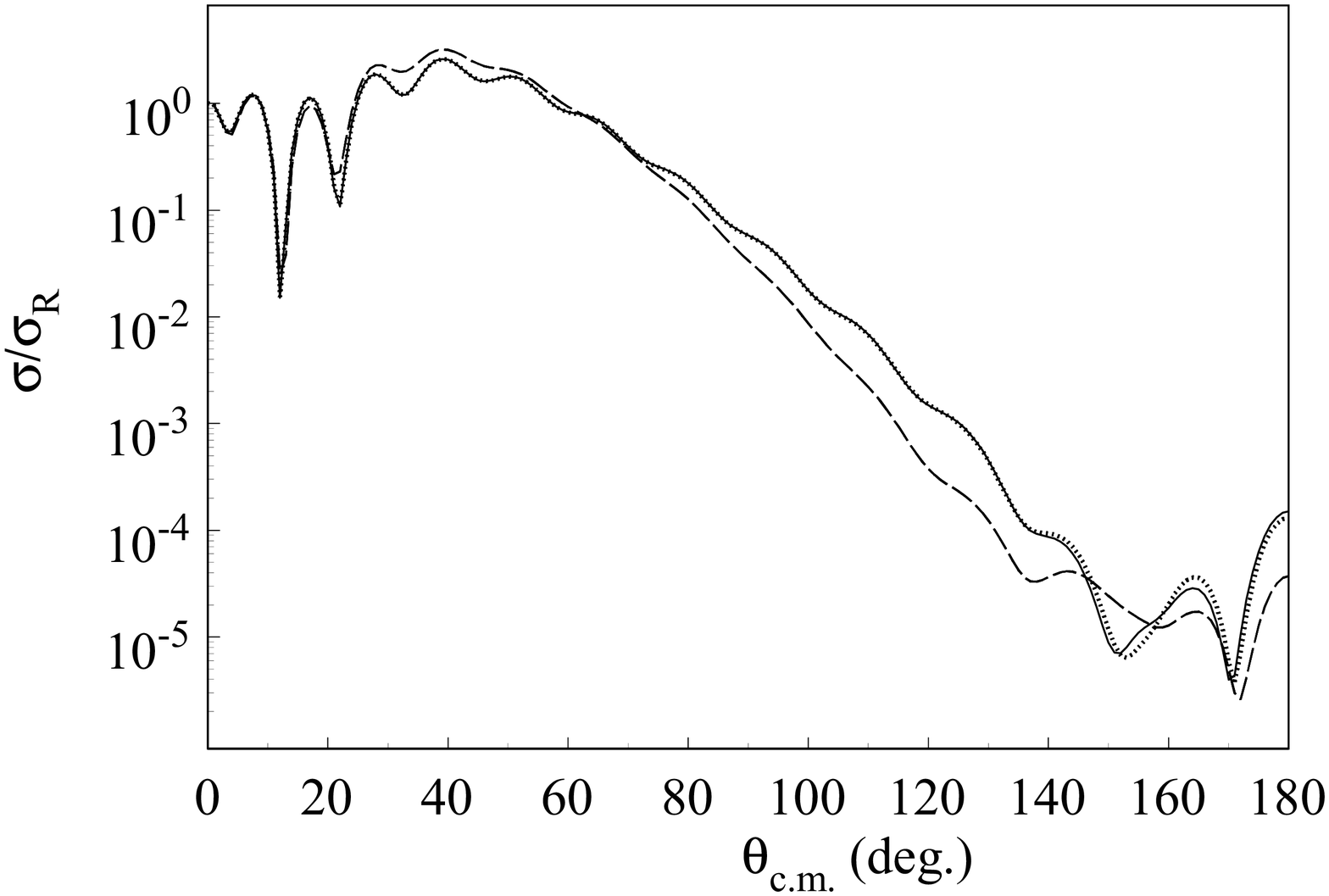}
\caption{\label{fig2} Comparison for the differential elastic
cross sections of the deuteron off $^{58}$Ni-target scattering (plotted as a ratio
ratio to the Rutherford cross section $\sigma_{\rm R}$) at $E_{\rm
lab}=80$ MeV calculated with the CDCC (dotted line) and WPCD
approaches (full line). The dashed curve corresponds to the
single folding model calculation.}
\end{figure}

At this relatively high energy, we found that the breakup
effect contribution is rather small, producing just a minor
correction to the folding model result (i.e.\ the calculation omitting continuum channels). Also it was shown
previously \cite{P} that only about half of the open channels (i.e.\  $E_{\rm
max}\approx 40$ MeV) play a significant role at this relatively high energy. So
one can conclude that the effect of  closed channels can be ignored
here. In Fig.~\ref{fig2} it is shown the comparison between the
elastic differential  cross sections obtained within the CDCC,  the
WPCD  and the folding model. For these calculations the WP
bases with dimensions $N=13$ (for n-p relative motion)\footnote{
 The Gaussian basis (\ref{gaub}) with $N_{G}=20$,
$t=2$, $\alpha_0=0.1$ fm$^{-1}$ has been used for this calculation.}
and $M=500$ (for deuteron c.m.\ motion) are used. The number of 
 partial waves required for the deuteron-target relative motion is $L_{\rm max}=45$. For the
CDCC calculations, the continuum was truncated at
$E_\mathrm{max}=70$~MeV and divided into $N=20$ bins, uniformly distributed in the linear momentum $k$. The set
of coupled equations were solved using the Numerov method, and
matched to their asymptotic solution at 20~fm. These calculations
were performed with the code FRESCO \cite{Th}.

As it follows from  Fig.~\ref{fig2} the results of both
discretization methods are in very good agreement for all angles.
It should be emphasized that despite the highly non-homogeneous bin
distributions for the internal $n-p$ subsystem as well as for the
deuteron-target subHamiltonian     lead  to  well
converged results of the WPCD calculations. So, it is not necessary to use in practical
calculations just equidistant momentum bin distributions  as it is
usually done \cite{R2,C1,P}.
 Moreover, the application of  some special energy distribution allows
to decrease significantly the number of the packet basis functions
required for convergence. Thus, to summarize, despite of the quite
different bin distributions considered in the CDCC and WPCD approaches
and rather different formalisms both methods provide results 
 in  nice agreement to each other at $E_{\rm lab}=80$ MeV.

\subsubsection{$E_{\rm lab}=$21.6  MeV. The effect of closed
channels}
 At this intermediate energy, the influence of 
closed channels (CC) on the elastic scattering amplitude becomes
visible. Our calculations have shown that the main part of this
effect is seen at backward angles. Though the CC effect has been
commonly neglected
 in the conventional
CDCC calculations\cite{J,R1,A,R2,F,L,C1,C2,P,C4} it is worth to
study carefully the convergence of the  elastic cross section when
the number of closed channels and the corresponding maximum energy
$E_{\rm max}$  are increased.

 To
prepare this calculations in the WPCD scheme we have used the
three-body WP sets of dimensions $N=15$ (constructed with the
Gaussian basis (\ref{gaub}) with  $N_{G}=30$,  $t=3$, $\alpha_0=0.1$
fm$^{-1}$) and $M=150$. The number of partial waves required  
at this energy is $L_{\rm max}=25$. To incorporate the closed
channels into the CDCC framework a special combination
 of the coupled-channel scheme and the R-matrix approach has been used 
 \cite{Th2}. In contrast to the Numerov method, where the radial functions
$\chi_i(\mathbf{R})$  are obtained by direct integration
of the coupled differential equations (\ref{CDCCeq}), 
in the R-matrix method a basis set of {\em energy eigenstates} is first obtained by
solving the diagonal parts of the coupled equations,
 with the basis functions all having fixed logarithmic
derivatives at a given distance $R_m$. Next,   the unknown functions $\chi_i(\mathbf{R})$
are expanded in this basis. Although this procedure is computationally more demanding, it leads
to more stable results when closed channels are taken into account. In order to achieve
convergence at this incident energy, we required $N_\mathrm{open}\simeq 20$ and
 $N_\mathrm{closed}\simeq 15$ bins, for a maximum excitation energy $E_\mathrm{max} \sim 140$~MeV. The
R-matrix radius was fixed to $R_m=30$~fm.

\begin{figure}
\includegraphics[width=9cm]{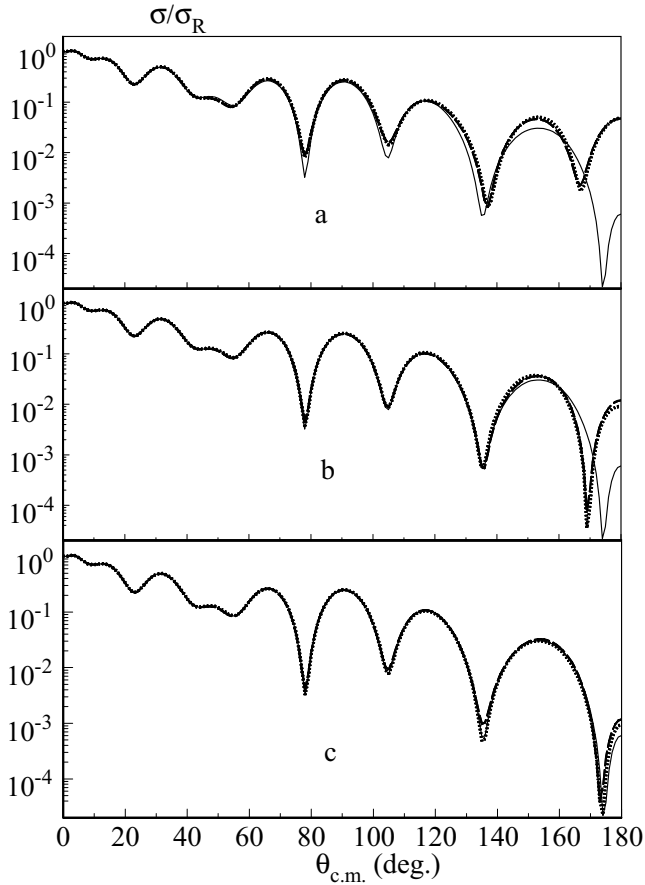}
\caption{\label{fig3} Elastic $d+^{58}$Ni cross section at $E_{\rm
lab}=21.6$ MeV calculated within the CDCC  (dotted line) and
WPCD (dashed line) approaches  for different maximum values
$E_{\rm max}$ of the excitation energy: a)  $E_{\rm max}=20$ MeV, b)
$E_{\rm max}=38$ MeV, c) $E_{\rm max}=80$ MeV. The full lines
correspond to the converged CDCC and WPCD results, which are almost
indistinguishable in these plots. }
\end{figure}

In  Fig.~\ref{fig3} the comparison between the R-matrix CDCC and the
WPCD elastic cross sections calculated at almost equal $E_{\rm max}$
values   is presented. For  
 $E_{\rm max}=20$ MeV, there are only open-channels and 
the conventional Numerov procedure has been employed to solve the CDCC equations (see 
Fig.~\ref{fig3}a). For the other
 values of $E_{\rm max}$ (Fig.~\ref{fig3}b and c), the R-matrix
 method has been employed, while the WPCD technique has
not been changed. It is evident that the results for the
differential cross sections found  with the WPCD and R-matrix CDCC
methods with similar $E_{\rm max}$ values are almost identical. So,
it can be concluded that the details of the discretization distributions in
both methods do not play an important role in the convergence
and the most significant factor here is just the $E_{\rm max}$ value.

\begin{figure}
\includegraphics[width=9cm]{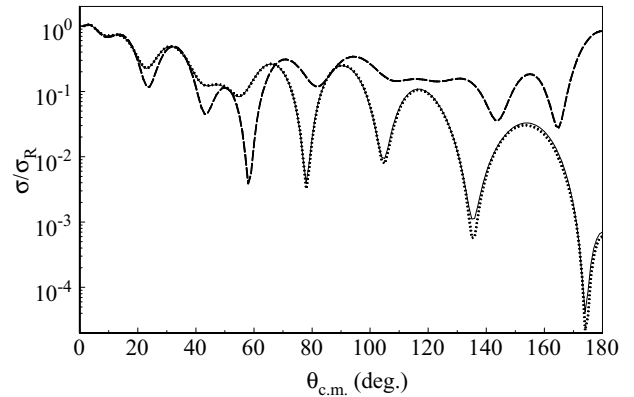}
\caption{\label{fig4} Converged results (with respect to the maximum energy $E_{\rm max}$) 
for the differential elastic cross section (relative to Rutherford) 
for $d+^{58}$Ni at $E_{\rm lab}=21.6$ MeV, calculated with the 
 CDCC (dotted curve) and  WPCD (solid curve) approaches. The dashed line corresponds to the 
folding calculation, in which continuum channels are not taken into account.}
\end{figure}

 The  converged elastic cross sections (calculated with the highest
 value, $E_{\rm max}=139$ MeV)
 are presented in  Fig.~\ref{fig4}. The curves corresponding to both discretization
  methods are
 hardly distinguishable. In this figure we have included also the calculation corresponding to
 the folding model calculation (dashed line). It is evident that breakup effects are large
 at this incident energy.

\begin{figure}
\includegraphics[width=9cm]{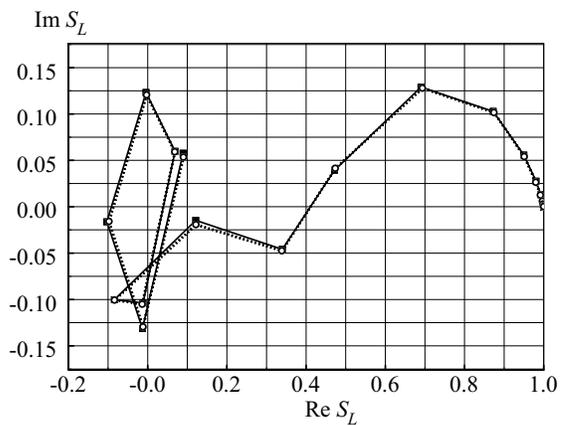}
\caption{\label{fig5} Argand plot for the elastic  $S$-matrix
elements for the reaction $d+^{58}$Ni at $E_{\rm lab}=21.6$ MeV calculated within the  CDCC  (open circles connected by dotted lines) 
 and WPCD  (black squares connected by solid lines) methods.}
\end{figure}

In Fig.~\ref{fig5} we plot the elastic $S$-matrix elements obtained with 
both discretization methods. It can be seen that they are in very good agreement for all partial waves. 

   From the present results it becomes clear that the CC effect must be taken into account
 at these intermediate energies. Furthermore, the required maximum  energy  of the CCs is rather
 high, viz.\ $E_{\rm max}\sim 100$ MeV.

\subsubsection{$E_{\rm lab}=$12  MeV}

 We finally present in this section the  calculations at the
relatively low energy of 12 MeV. In this case
the  dimensions of the WP sets  $N=15$ (constructed from the
Gaussian basis (\ref{gaub}) with $N_{G}=30$, $t=3$, $\alpha_0=0.1$
fm$^{-2}$) and $M=150$ are sufficient, and the maximum deuteron
c.m.\ angular momentum value is only $L_{\rm max}=15$. For the CDCC
calculations we used $N_\mathrm{open}= 15$ and
$N_\mathrm{closed}=10$ (for $E_\mathrm{max} \sim 80$~MeV). Again,
the R-matrix technique was used to solve the coupled equations, with
a matching radius of $R_m=30$~fm. 

\begin{figure}
\includegraphics[width=9cm]{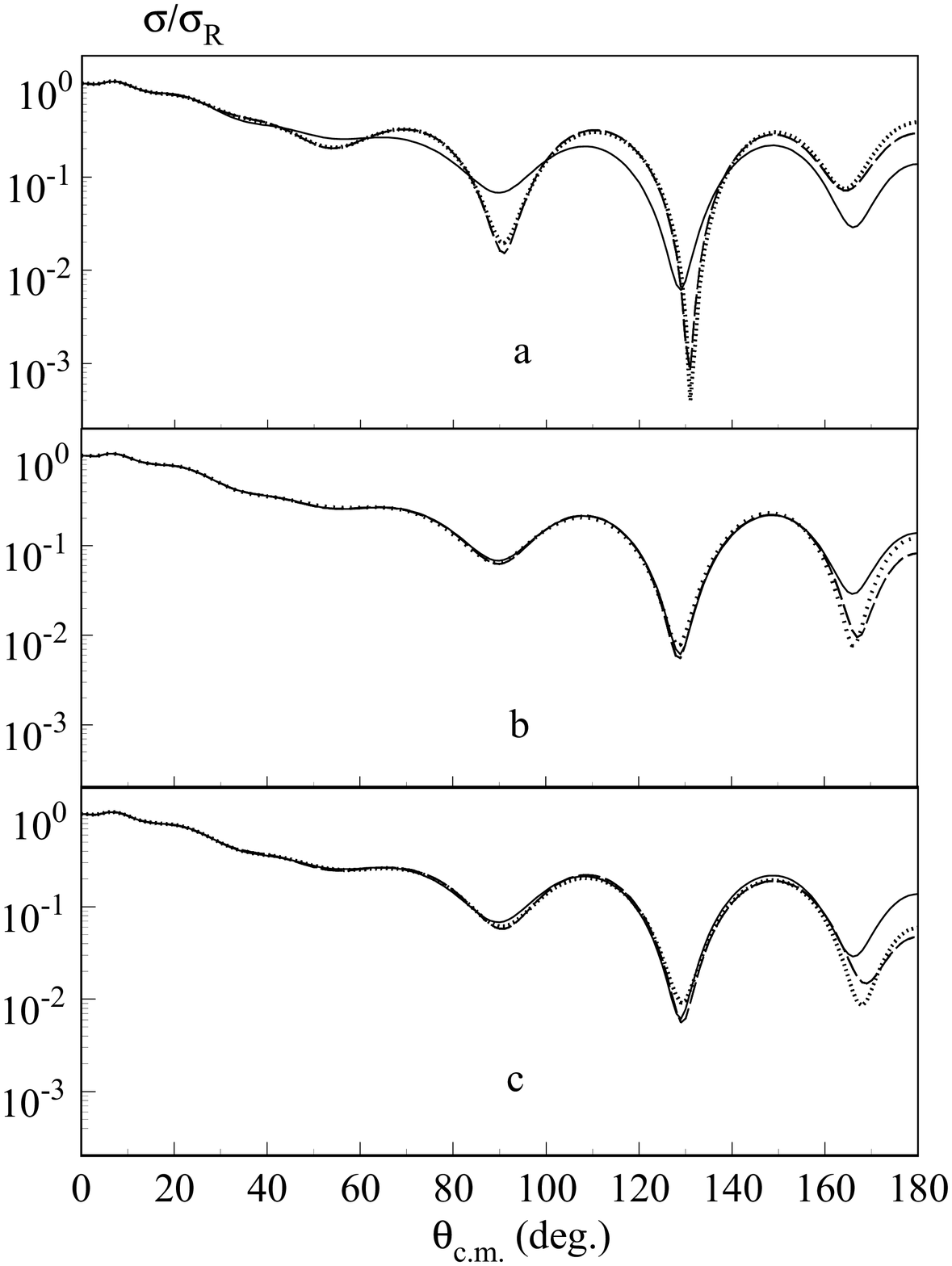}
\caption{\label{fig6} Elastic $d+^{58}$Ni cross section at $E_{\rm
lab}=12$ MeV calculated within the   CDCC (dotted line) and
WPCD (dashed line) approaches for different maximum values of the 
excitation  energy, $E_{\rm max}$: a)  $E_{\rm max}=10.2$ MeV, b)
$E_{\rm max}=24.5$ MeV, c)  $E_{\rm max}=36$ MeV. The full curves
correspond to the converged CDCC results.}
\end{figure}

\begin{figure}
\includegraphics[width=9cm]{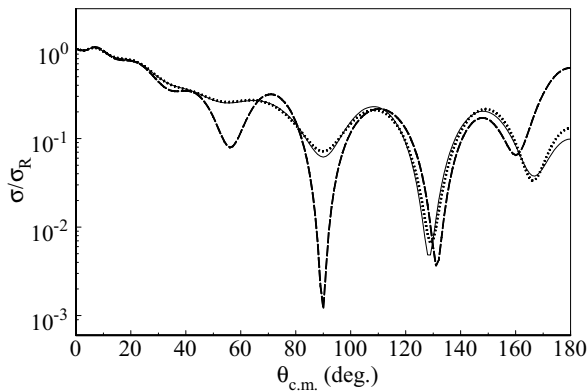}
\caption{\label{fig7} The same as in Fig.~\ref{fig4}, but  
for $E_{\rm lab}=12$ MeV.}
\end{figure}

 The elastic cross sections calculated with  the
WPCD and the R-matrix CDCC approaches for different $E_{\rm max}$
values  are shown in  Figs.~\ref{fig6}a, b and c.  The comparison
between the converged (with respect to $E_{\rm max}$ values) cross
sections calculated with both methods along with the folding model
results are presented in Fig.~\ref{fig7}.

It is evident that the influence of the CC is the largest here and this effect  is
seen not only at backward angles but also at intermediate angles
($\theta_\mathrm{c.m.} \sim 90^\circ-100^\circ$). The small discrepancies
between the WPCD and R-matrix CDCC results observed 
at most backward angles  might be ascribed
 to the different n-p continuum discretization
parameters, but we can not rule out that they can be related 
 to numerical inaccuracies of either of the two methods. In any case, this 
effect, which has not been seen in the two previous
cases, begins to play some role at rather low incident energies.

\section{Discussion}
The close agreement between the WPCD and R-matrix CDCC approaches
found in the present work can be interpreted as a good indicator
that the basic assumptions (or approximations) adopted in the WPCD
method, in particular the employment of orthonormalized stationary
wave packets instead of exact non-normalizable three-body scattering
wave functions are well justified and correct. This fact implies
some further consequences in the whole many-body scattering
theory. Actually, the usage of the total continuum discretization on
the basis of normalizable (of $L_2$ type) stationary wave packets is
equivalent to the formulation of the many-body scattering problem in
a finite multi-dimensional box and leads a treatment of
the original scattering problem  analogous to the much more simple bound-state
problem. In  turn, the last conclusion allows to treat the
many-body scattering problem in  similar way as the many-body bound
states, e.g.\ by expanding the $L_2$-type stationary wave packets
(constructed from the exact unknown scattering wave functions) on some
convenient $L_2$ basis widely used in bound state problems (for
example, the many-body harmonic oscillator basis). The first
results which two of the present authors \cite{K2} have obtained with the
 application of this technique for the solution of the Faddeev equations above
the three-body breakup threshold seems to support this expectation.

Another nice feature of the WPCD approach is the possibility 
to  reformulate in a straightforward way the whole problem in terms of the
Feshbach projection technique \cite{K4} and to include all the
breakup channels into the respective Feshbach polarization non-local
potential. After this, one can readily use the full
applicability of the WPCD approach to the resulted non-local
interactions (in contrast to the conventional coupled-channel
method) and to reduce in a standard way the complicated many-body
scattering problem to the simple matrix equations \cite{new}.

On the other hand, the very close agreement between the conventional
CDCC approach and the fully alternative wave-packet integral method
should remove any doubts reported in literature
\cite{Sawada,Red} concerning the reliability and convergence properties
of the Schroedinger coupled-channel scheme in the framework of the 
three-body scattering problem.

\section{Conclusion}
In this work, we have compared two essentially different approaches based
on the general idea of continuum discretization for solving
three-body scattering problems. The CDCC method was proposed to
solve the coupled-channel problem in the framework of the
Schroedinger equation formalism. This approach uses a discretization 
of the projectile internal continuum only, 
along with
an explicit matching between inner and asymptotic scattering wave
functions to extract the multichannel $S$-matrix. On the other hand, 
the WPCD approach
is based on the total three-body continuum discretization which
allows to use the integral equation formalism of the scattering
theory. The special wave-packet technique allows to construct
finite-dimensional analogs for all the scattering operators and to
use only inner parts of wave functions in order to find the
observables.

The direct comparison of both methods  leads to the following
conclusions:
\begin{itemize}
\item[(i)]
 {The two methods of  continuum
discretization discussed in this work produce very close results for the composite particle
elastic scattering at rather high and also at low energies.}

\item[(ii)]{Different types of discretization of the
 projectile internal subHamiltonian continuum used in both methods do not lead
 to visible discrepancies at high and intermediate energies.
  Small differences in the elastic cross sections arise only at low incident energies.
  An employment of a highly non-homogeneous dicretization distribution in the
   WPCD approach still leads to well-converged  results.}

\item[(iii)]{ The effect of closed channels
is not seen at high incident energies. This effect  begins
to play a visible role at intermediate energies (at the most backward
angles)
 and can not be ignored
as it was assumed in some previous CDCC calculations. The
convergence of the calculated cross section when increasing  the number of
included CCs depends only
on the maximum excitation energy  and is not sensitive
to the details of the continuum dicretization.}

\item[(iv)]{  The effect of CC on the elastic  scattering becomes more
significant when the incident energy of the projectile  decreases.}

\item[(v)] {The very close agreement between the results obtained with 
these two different approaches at various energies, from low to high, seems to
indicate that the three-body wave-packet technique can be considered
as a rather reliable and numerically accurate method for calculations
of composite projectile scattering by nuclei. Simultaneously, one
can confirm again that the conventional (or R-matrix) CDCC method
leads to quite reliable and well-converged results for the elastic
and inelastic scattering of composite projectiles.}
\end{itemize}

{\bf Acknowledgements} The Russian authors (V.I.K.\ and O.A.R.) are
thankful to Dr.\  V.N.\ Pomerantsev for numerous fruitful discussions. They appreciate very much
  partial financial supports from the RFBR grant 07-02-00609,
the joint RFBR--DFG grant 08-02-91959 and the President grant
MK-202.2008.2.
 A.M.M.\ acknowledges a research grant from the Junta de Andaluc\'{\i}a
and financial support by the DGICYT under project FPA2006-13807-c02-01.

\end{document}